\documentclass[npg]{copernicus}

\hypersetup{draft}

\begin{document}


\title{\vspace{-6.08em}
On~the~CCN~[de]activation~nonlinearities}

\Author[1,2]{Sylwester}{Arabas}
\Author[3]{Shin-ichiro}{Shima}

\affil[1]{Institute of Geophysics, Faculty of Physics, University of Warsaw, Warsaw, Poland}
\affil[2]{Chatham Financial Corporation Europe, Cracow, Poland}
\affil[3]{Graduate School of Simulation Studies, University of Hyogo, K\=obe, Japan}

\runningtitle{On the CCN [de]activation nonlinearities}

\runningauthor{Arabas \& Shima}

\correspondence{Sylwester Arabas (\url{sarabas@chathamfinancial.eu}) and Shin-ichiro Shima (\url{s_shima@sim.u-hyogo.ac.jp})}

\received{}
\pubdiscuss{}
\revised{}
\accepted{}
\published{}

\firstpage{1}
\maketitle

\begin{abstract}
We take into consideration the evolution of particle size in a monodisperse 
  aerosol population during activation and deactivation of cloud 
  condensation nuclei (CCN).
The phase portrait of the system derived through a 
  weakly-nonlinear analysis reveals a saddle-node bifurcation
  and a cusp catastrophe.
An analytical estimate of the activation timescale is derived through
  estimation of the time spent in the saddle-node bifurcation bottleneck.
Numerical integration of the system portrays two types of 
  activation/deactivation hystereses: 
  one associated with the kinetic limitations on droplet growth 
  when the system is far from equilibrium, and 
  one occurring close to equilibrium and associated with
  the cusp catastrophe.
The hysteretic behaviour close to equilibrium imposes stringent time-resolution 
  constraints on numerical integration, particularly during deactivation.

\end{abstract}

\vspace{-1.6em}
\introduction[Background]
\vspace{-.5em}

Atmospheric clouds are visible to human eye for they are composed of water 
  and ice particles that effectively scatter solar radiation.
The multi-micrometre light-scattering cloud droplets form on sub-micrometre 
  aerosol particles
  (cloud condensation nuclei, CCN) in a process referred to as CCN activation.
The concentration (100--1000 cm\textsuperscript{-3}) and mean size (1--10 $\mu m$) of 
  activated particles can vary by an order of magnitude depending on the size
  spectrum and composition of CCN.
On one hand, CCN physicochemical properties are influenced by anthropogenic 
  emissions of particles into the atmosphere.
On the other hand, the resultant size spectrum of cloud droplets determines how 
  effectively the clouds interact with solar radiation and how effectively
  they produce precipitation
  \citep[see e.g. a recent NPG paper by][for a discussion of aerosol-cloud-precipitation interaction 
  chain, unconventionally modelled as a predator-prey problem]{Feingold_and_Koren_2013}.
CCN~activation is thus the linking process
  between the microscopic human-alterable atmospheric composition 
  and the macroscopic climate-relevant cloud properties.
As once aptly stated, ``{\it there is something captivating about the idea that 
  fine particulate matter, suspended almost invisibly in the atmosphere, holds 
  the key to some of the greatest mysteries of climate 
  science}''~\citep{Stevens_2012}.
This has certainly contributed to the wealth of literature on the subject
  published since the first studies of the 1940-ties
  \citep{Howell_1949,Tsuji_1950}, 
  for a thorough list of references see e.g., \citet[chpt.~7]{Khvorostyanov_and_Curry_2014}.

Deactivation is the reverse process in which cloud droplets evaporate
  back to aerosol-sized particles.
The process is also referred to as aerosol regeneration, 
  aerosol recycling, drop-to-particle conversion or simply droplet evaporation
  \citep[see section~1 in][for a review of modelling studies]{Lebo_and_Seinfeld_2011}.
Both activation and deactivation are particular cases of particle 
  condensational growth which, in context of cloud modelling, is generally 
  regarded as reversible to contrast the irreversible collisional growth
  \citep[see e.g.,][]{Grabowski_and_Wang_2013}.
The reversibility of condensational growth is a sound (and often a constituting) 
  assumption for cloud models
  for which activation and deactivation are subgrid processes, both in terms of
  time- and length-scales.
Yet, when investigated in short-enough timescales, 
  condensation and evaporation exhibit a hysteretic behaviour 
  in an activation-deactivation cycle.
The hysteresis can be associated with the kinetic limitations
  in the vapour and heat transfers to/from the droplets \citep{Chuang_et_al_1997}
  and has been previously depicted in the studies of 
  \citet[discussion of Fig.~1]{Korolev_2003}
  and \citet{Korolev_et_al_2013}.
As we point out in this note, the system can exhibit a hysteretic behaviour 
  also in a close-to-equilibrium r\'egime where the kinetic 
  limitations do not play a significant role.

\vspace{-1.6em}
\section{Droplet growth laws in a nutshell}
\vspace{-.5em}

\begin{figure}[t]
  \center
  \includegraphics[width=.45\textwidth]{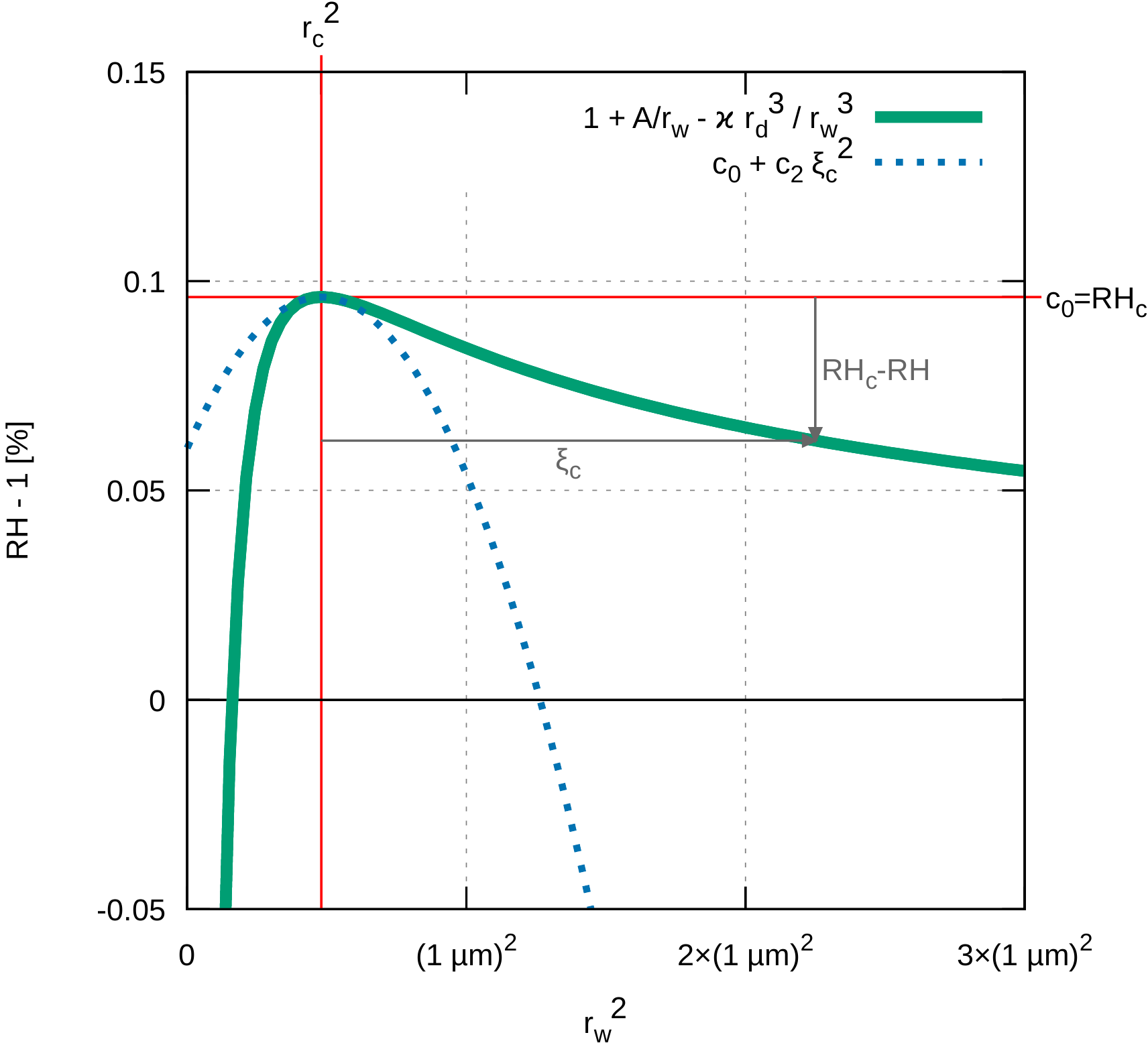}
  \caption{\label{fig:koehler}
    K\"ohler curve for CCN with $r_\text{d}=0.05 \mu m$, $\kappa=1.28$ (\chem{NaCl}) 
    and its Taylor expansions at $r_\text{c}$ and at infinity.
  }
\end{figure}

The key element in the mathematical description of 
  CCN activation/deactivation is the equation for the 
  rate of change of particle radius $r_\text{w}$ (so-called wet radius)
  due to water vapour transfer to/away from the particles. 
It is modelled by a diffusion equation in a spherical geometry:
\begin{equation}
  \dot{r}_\text{w} 
  = \frac{1}{r_\text{w}} \frac{D_{\text{eff}}}{\rho_\text{w}} \left( \rho_\text{v} - \rho_\circ\right) 
\end{equation}
  where
  $\rho_\text{w}$~is the liquid water density, $\rho_\text{v}$ is the ambient
  vapour density (away from the droplet), $\rho_\circ$~is the equilibrium
  vapour density at the drop surface and the 
  $D_\text{eff}=D_\text{eff}(T,r_\text{w})$ is an effective diffusion coefficient
  in which the temperature dependence stems from an approximate combination of the Fick’s 
  first law and Fourier’s law (latent heat release) into a single particle-growth equation
  (so-called Maxwell-Mason formula), while the radius dependence stems from 
  corrections limiting the diffusion efficiency for smallest particles.
For derivation and discussion see \citet[section 5.1.4]{Khvorostyanov_and_Curry_2014}.
Introducing two non-dimensional numbers: 
  the relative humidity $\text{RH}=\rho_\text{v}/\rho_\text{vs}$
  (the ratio of the ambient vapour density to the vapour density at saturation 
  with respect to plane surface of pure water) and the equilibrium relative
  humidity $\text{RH}_\text{eq}=\rho_{\circ}/\rho_\text{vs}$, the drop growth
  equation is given by:
\begin{equation}\label{eq:dg_rh}
  \dot{r}_\text{w} = \frac{1}{r_\text{w}} D_{\text{eff}} \frac{\rho_\text{vs}}{\rho_\text{w}} \left(\text{RH} - \text{RH}_\text{eq}\right)
\end{equation}

The crux of the matter is the dependence of $\text{RH}_\text{eq}$ on $r_\text{w}$
  due to the droplet curvature and due to the presence of dissolved substances.
It is given by the so-called K\"ohler curve, the leading terms of
  its common $\kappa$-K\"ohler form can be approximated with (for 
  $r_\text{d} \ll r_\text{w}$ what is an reasonable assumption in context
  of activation/deactivation):
\begin{eqnarray}
  \text{RH}_\text{eq} 
  &\!\!=\!\!& 
  \frac{r_w^3 - r_d^3}{r_w^3 - r_d^3 (1 - \kappa)}
  \exp\!\left(\frac{A}{r_w}\right)
  \\\label{eq:koehlerapprox}
  &\!\!\approx\!\!&
  1 + \frac{A}{r_\text{w}} - \frac{\kappa r_\text{d}^3}{r_\text{w}^3} 
\end{eqnarray}
  where $A\sim10^{-3}\mu m$ is a temperature-dependant coefficient
  related with surface tension of water, while the dry radius $r_\text{d}$ 
  and the solubility parameter $\kappa$ 
  \citep[in general, $0<\kappa<1.4$, see][]{Petters_and_Kreidenweis_2007} are proxy 
  variables depicting the mass and chemical composition of the substance the 
  CCN are composed of.
The $\frac{\partial}{\partial_{r_\text{w}}} \text{RH}_\text{eq}$ 
  derivative has an analytically-derivable root corresponding to the maximum
  of the K\"ohler curve at ($r_\text{c}, \text{RH}_\text{c}$) where 
  $r_\text{c} = \sqrt{3 \kappa r_\text{d}^3 / A}$ is the so-called critical radius
  and $\text{RH}_\text{c}=1 + \frac{2A}{3 r_\text{c}}$ is the critical relative humidity.
Neglecting the details of the activation kinetics, the CCN 
  can be considered activated (deactivated) when they grow beyond 
  (shrink below) their critical radius.

\vspace{-1.6em}
\section{Saddle-node bifurcation at K\"ohler curve maximum}\label{sec:saddle}
\vspace{-.5em}

Rewriting equation~\ref{eq:dg_rh} in terms of $\xi = r_\text{w}^2 + C$ (where $C$ is an arbitrary constant) gives:
\begin{equation}\label{eq:xi_dg}
  \dot{\xi} = 2 D_{\text{eff}} \frac{\rho_\text{vs}}{\rho_\text{w}} \left( \text{RH} - \text{RH}_\text{eq}\right)
\end{equation}  
Rewriting $\text{RH}_\text{eq}$ in terms of $\xi_\text{c}=r_\text{w}^2 - r_\text{c}^2$ and Taylor-expanding it around $\xi_\text{c}=0$ gives:
\begin{equation}\label{eq:xi_kc}
  \text{RH}_\text{eq}(\xi_\text{c}) = c_0 + \cancel{c_1 \xi_\text{c}} + c_2 \xi_\text{c}^2 + \ldots
\end{equation}
  where $c_0=\text{RH}_\text{c}$, $c_1$ is zero as we are expanding around the root of
  $\partial_{\xi_\text{c}} \text{RH}_\text{eq}$ and $c_2=-\frac{A}{4}r_\text{c}^{-5}$ is negative. 
Combining equations~\ref{eq:xi_dg}~and~\ref{eq:xi_kc} gives:
\begin{equation}\label{eq:pp_r_c}
  \left. \dot{\xi_\text{c}} \right|_{\xi_\text{c}\rightarrow 0} \sim \frac{\text{RH} - \text{RH}_\text{c}}{A / (4 r_\text{c}^5)} + \xi_\text{c}^2 
\end{equation}
  which is the normal form of the saddle-node bifurcation 
  \citep[section~3.1]{Strogatz_2014}.
The phase portrait 
  of the system can be recognised in a standard cloud-physics K\"ohler curve 
  plot given in Figure~\ref{fig:koehler}.
In the nearest vicinity of the critical radius, when $\text{RH}$ approaches $\text{RH}_\text{c}$
  from below, there are two fixed points in the system: 
  one stable (for $r_\text{w} < r_\text{c}$, unactivated CCN) and 
  one unstable (for $r_\text{w} > r_\text{c}$, activated CCN).
The bifurcation occurs when $\text{RH}=\text{RH}_\text{c}$ when the fixed points coalesce
  into a half-stable fixed point.
When $\text{RH}>\text{RH}_\text{c}$, there are no fixed points at all.

\vspace{-1.6em}
\section{Activation timescale estimation}\label{sec:timescale}
\vspace{-.5em}

\begin{figure}[t]
  \includegraphics[width=.45\textwidth]{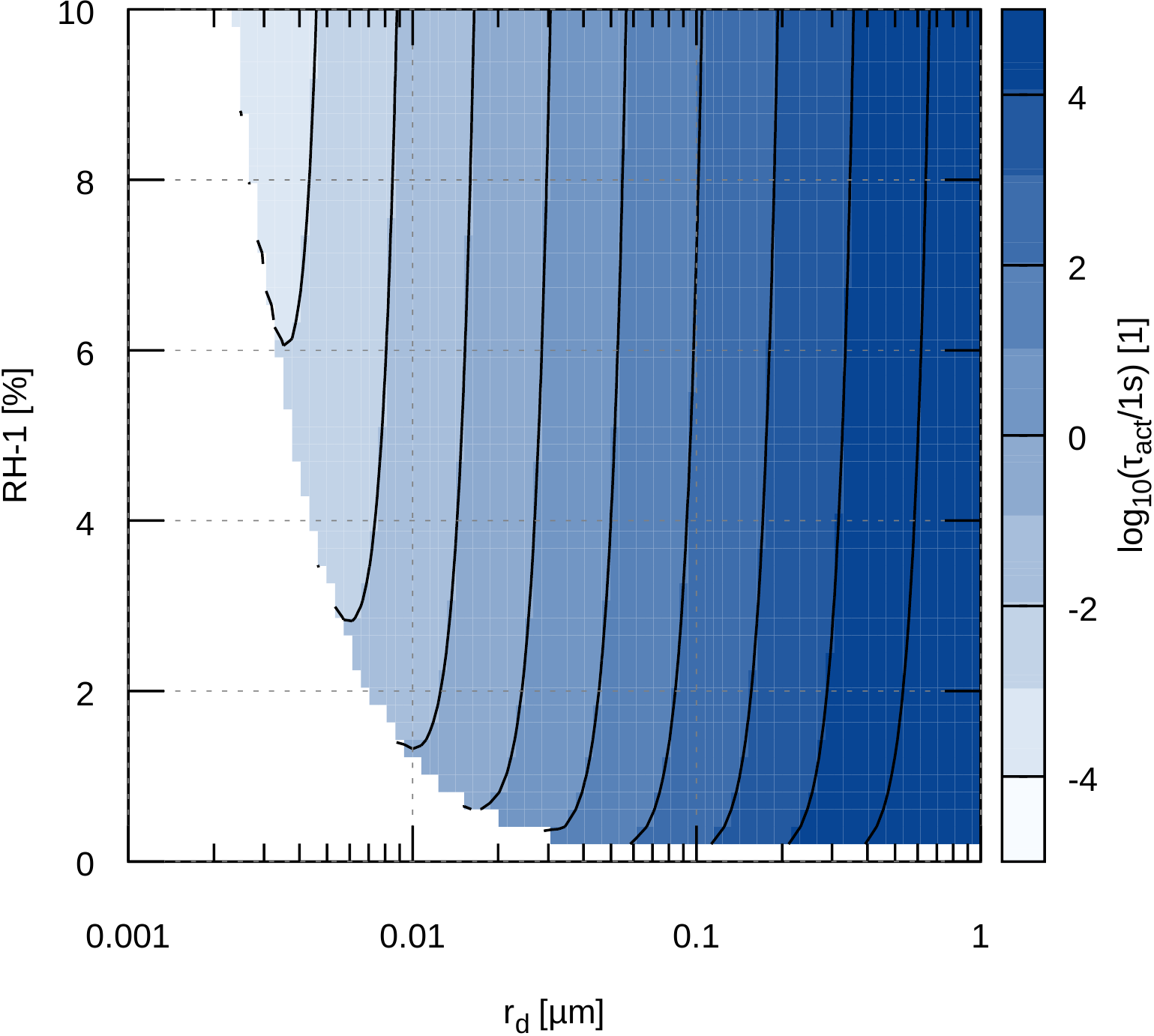}
  \caption{\label{fig:tau_act}
    Activation timescale as a function of dry radius and relative 
      humidity estimated with equation \ref{eq:tau} with $A\sim10^{-3}\mu m$,
      $\kappa=1.28$, $D\sim2\times10^{-5} \frac{m^2}{s}$, $\rho_\text{w}\sim10^3 \frac{kg}{m^3}$
      and $\rho_\text{vs}=10^{-3} \frac{kg}{m^3}$.
  }
\end{figure}

Interestingly, the analysis of the CCN activation/deactivation in terms
  of saddle-node bifurcation provides a way to estimate the
  timescale of activation.
Following \citet[section~4.3]{Strogatz_2014}, 
  the coalescence of the fixed points is associated with a passage through 
  a {\em bottleneck}.
The key observation is that for the parabolic normal form of the saddle-node 
  bifurcation, the time of the passage through the bottleneck dominates all
  other timescales.
Thus, the timescale of the process can be estimated by integrating $\xi_\text{c}$ 
  from~$-\infty$~to~$\infty$:
\begin{equation}\label{eq:tau}
  \tau_{act} 
    \approx \int_{-\infty}^{+\infty} \frac{d\xi_\text{c}}{\dot{\xi_\text{c}}} 
    = \frac{r_c^{5/2}}{\sqrt{A}}\frac{\rho_\text{w}/\rho_\text{vs}}{D_\text{eff}} \frac{\pi}{\sqrt{\text{RH}-\text{RH}_\text{c}}}\\
\end{equation}
The activation timescale $\tau_\text{act}$ given by eq.~\ref{eq:tau}, plotted
  as a function of $\text{RH}$ and $r_\text{d}$ (and substituting $r_\text{c}$ and 
  $\text{RH}_\text{c}$ by their analytic formulae given in the preceding section) 
  is presented in Figure~\ref{fig:tau_act}.
It matches remarkably the data obtained through numerical calculations presented in 
  \citet{Hoffmann_2016}.
The white region in the plot corresponds to situation where activation does not 
  happen.
The range of $\text{RH}$ depicted in the plot is chosen to match the one of~Figure~2~in
  \citet{Hoffmann_2016}, while in principle the presented weakly non-linear 
  analysis of the system is applicable only close to the equilibrium (i.e.,
  close to the edge of the white region in the plot).

\vspace{-1.6em}
\section{Cusp catastrophe of the RH-coupled system}\label{sec:cusp}
\vspace{-.2em}

The key limitation of the preceding analysis is that the evolution of particle
  size is not coupled with the evolution of ambient heat and moisture content,
  and hence the relative humidity.
Limiting the analysis to a monodisperse population, the coupling efficiency 
  is determined by the total number of particles in the system.
The so-far assumed constant RH approximates thus the case of small number of droplets.

To at least partially lift the constant-RH assumption, while still allowing
  for a concise analytic description of the system dynamics,
  let us consider a simple representation of the moisture budget in the system
  under a temporary assumption of constant temperature and pressure 
  (and hence constant volume, constant $\rho_\text{vs}$, $A$ and $D_\text{eff}$).
The rate of change of the ambient relative humidity $\dot{\text{RH}}$
  can be expressed then as a function of the droplet volume concentration $N$:
\begin{equation}
  \dot{\text{RH}} \approx \frac{\dot{\rho}_\text{v}}{\rho_\text{vs}} = - N \underbrace{\frac{4\pi \rho_\text{w}}{3\rho_\text{vs}}}_{\alpha} 3 r_\text{w}^2 \dot{r}_\text{w} 
\end{equation}
where the form of $\alpha$ stems from defining the density of liquid water
  in the system as $N \rho_\text{w} \frac{4}{3} \pi r_\text{w}^3$.
Integrating in time gives:
\begin{equation}
  \text{RH} = \text{RH}_0 - \alpha N r_\text{w}^3
\end{equation}
which combined with eq.~\ref{eq:dg_rh} and expressed in terms of $\xi$ with $C=0$
  leads to the following phase portrait of the RH-coupled system (assuming $r_w \gg r_d$):
\begin{equation}\label{eq:f}
  \dot{\xi} \sim (\text{RH}_0-1) - \underbrace{\left(\frac{A}{\xi^{\frac{1}{2}}} - \frac{\kappa r_\text{d}^3}{\xi^{\frac{3}{2}}} + \alpha N \xi^\frac{3}{2} \right) }_f
\end{equation}
where the group of terms labelled as $f$ can be intuitively
  thought of as corresponding to the K\"ohler curve with an additional term 
  representing the simplified RH dynamics.

\begin{figure}[t]
  \center
  \includegraphics[width=.48\textwidth]{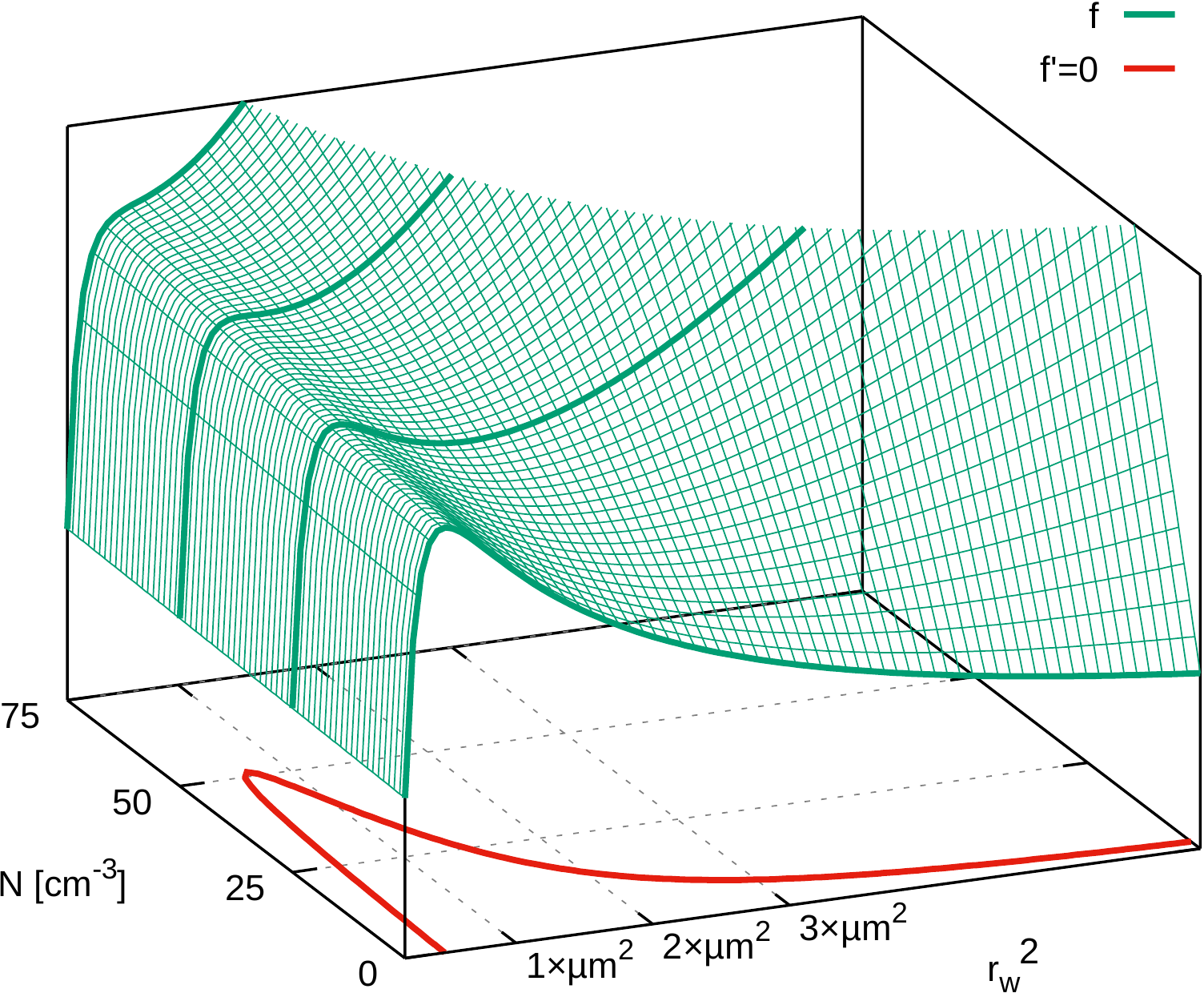}
  \caption{\label{fig:cusp}
     Dependence of $f$ defined in eq.~\ref{eq:f} on the wet radius
       and particle concentration (green wireframe surface).
     Red line below depicts the zero-crossings of the first derivative of $f$
       with respect to $r_w$.
     Values of all constants as in Fig.~\ref{fig:tau_act}.
     Discussion in section~\ref{sec:cusp}.
  }
\end{figure}

Figure~\ref{fig:cusp} depicts the dependence of $f$ on the droplet 
  radius $r_w=\sqrt{\xi}$ and the droplet concentration $N$.
To facilitate analysis, the zero-crossings of the first derivative of $f$ with respect
  to $r_w$ are plotted as well using analytically derived formula:
\begin{equation}
  {\rm sgn}(f') = {\rm sgn}\left(\kappa r_d^3 -\frac{A}{3} r_w + \alpha N r_w^3 \right)
\end{equation}

For $N=0$, $f$ has the K\"ohler-curve shape depicted in Fig.~\ref{fig:koehler}
  which, as discussed in the preceding sections, implies a saddle-node bifurcation.
With $N$ greater than zero but less than ca. 50/cm\textsuperscript{3}, a second saddle-node 
  point appears as the $\alpha N \xi^{\frac{3}{2}}$ term causes $f$ to 
  have a local minimum above the critical radius.
At ca. N=50/cm\textsuperscript{3}, both the first and second derivatives of $f$ vanish
  implying a cusp point in the $f$ surface.
For larger $N$, $f$ is monotonic, hence the saddle-node bifurcation ceases
  to exist in the system.
This phase portrait reveals a cusp bifurcation
  with the plotted zero-crossing line depicting its stability diagram.
The cusp bifurcation \citep[chpt.~8.2]{Kuznetsov_2004}, an imperfect supercritical
  pitchfork bifurcation \citep[chpt.~3.6]{Strogatz_2014},
  features a cusp catastrophe what allows to envision
  a ''catastrophic'' jump from one equilibrium to another and
  a hysteretic behaviour of the system when approaching (in therm of $r_w$) the
  local minimum of $f$ from below (activation) and from
  above (deactivation) for small enough $N$.

\vspace{-.6em}
\section{Adiabatic vertically-displaced air parcel system}\label{sec:parcel}
\vspace{-.5em}

In order to lift the assumptions of constant temperature and pressure, 
  the system evolution can be formulated by supplementing the drop growth
  equation with two equations representing the hydrostatic balance and
  the adiabatic heat budget.
This leads to a commonly used so-called air-parcel 
  framework depicting behaviour of a vertically displaced adiabatically
  isolated mass of air:
\begin{equation}\label{eq:parcel}
  \frac{d}{dt} \!\! \left[\!\!\begin{array}{c} 
    p_d\\
    T\\
    r_w
  \end{array}\!\!\right] 
  =
  \left[\begin{array}{c}
    -\rho_\text{d} g w\\
    ( \dot{p}_\text{d} / \rho_\text{d} - \dot{q} l_\text{v} ) / c_\text{pd} \\
    \text{(Eq.~1)}
  \end{array}\right]
\end{equation}
where $\rho_\text{d}$ and $p_\text{d}$ are the dry-air (background state)
  density and pressure, 
  $w$ is the vertical velocity of the parcel,
  $q=\rho_v / \rho_d$ is the water vapour mixing ratio, 
  $c_\text{pd}$ is the specific heat of dry air,
  $l_v$ is the latent heat of vapourisation and
  $g$ is the acceleration due to gravity.

As discussed in section~\ref{sec:cusp}, for 
  a monodisperse population of $N$ particles, the
  changes in the mass of liquid water in the system 
  are proportional to the particle concentration, hence $\dot{q}\sim N$.
Consequently, the analysis of the activation/deactivation dynamics presented in 
  sections~\ref{sec:saddle}-\ref{sec:timescale} under the assumption of constant $\text{RH}$ 
  corresponds to the behaviour of the air-parcel system defined by eq.~\ref{eq:parcel} in the limit of:
\begin{itemize}
  \vspace{-.2em}
  \item{
    $w\rightarrow 0$ (and hence $\dot{p}_\text{d}\approx0$)
    i.e., slow, close-to-equilibrium evolution of the system relevant to 
    fixed-point analysis
    (by some means pertinent to the formation of non-convective clouds such as fog) and
  }
  \vspace{-.2em}
  \item{
    $N\rightarrow 0$ (and hence $\dot{r}\approx 0$)
    i.e., weak coupling between particle size evolution and the ambient 
    thermodynamics (pertinent to the case of low particle concentration).
  }
  \vspace{-.2em}
\end{itemize}

\vspace{-1.6em}
\section{Numerical simulations}\label{sec:rhloop}
\vspace{-.5em}

Since the system defined by (\ref{eq:parcel}) is less susceptible to a simple
  analytic analysis, we proceed with numerical integration.
Furthermore, employing numerical integration allows to evaluate the K\"ohler
  curve in unapproximated form~(\ref{eq:koehlerapprox}) to corroborate
  the findings obtained with the assumption of $r_\text{d} \ll r_\text{w}$.
To this end, a numerical solver was implemented using the {\em libcloudph++} library 
  \citep{Arabas_et_al_2015}
  and the CVODE adaptive-timestep integrator
  \citep{Hindmarsh_et_al_2005}.
The solver code is free and open-source and is available as an
  electronic supplement to this note.

In order to depict an activation-deactivation cycle, the vertical velocity $w$
  was set to a sinusoidal function of time~$t$ such that the maximal
  displacement is reached at $t=t_\text{hlf}$ and the average velocity is $<\!\!w\!\!>$:
\begin{equation}
  w = \, <\!\!w\!\!> \frac{\pi}{2} \, {\rm sin}\!\left(\pi \frac{t}{t_\text{hlf}}\right)
\end{equation}

\begin{figure*}
  \center
  \includegraphics[width=.9\textwidth]{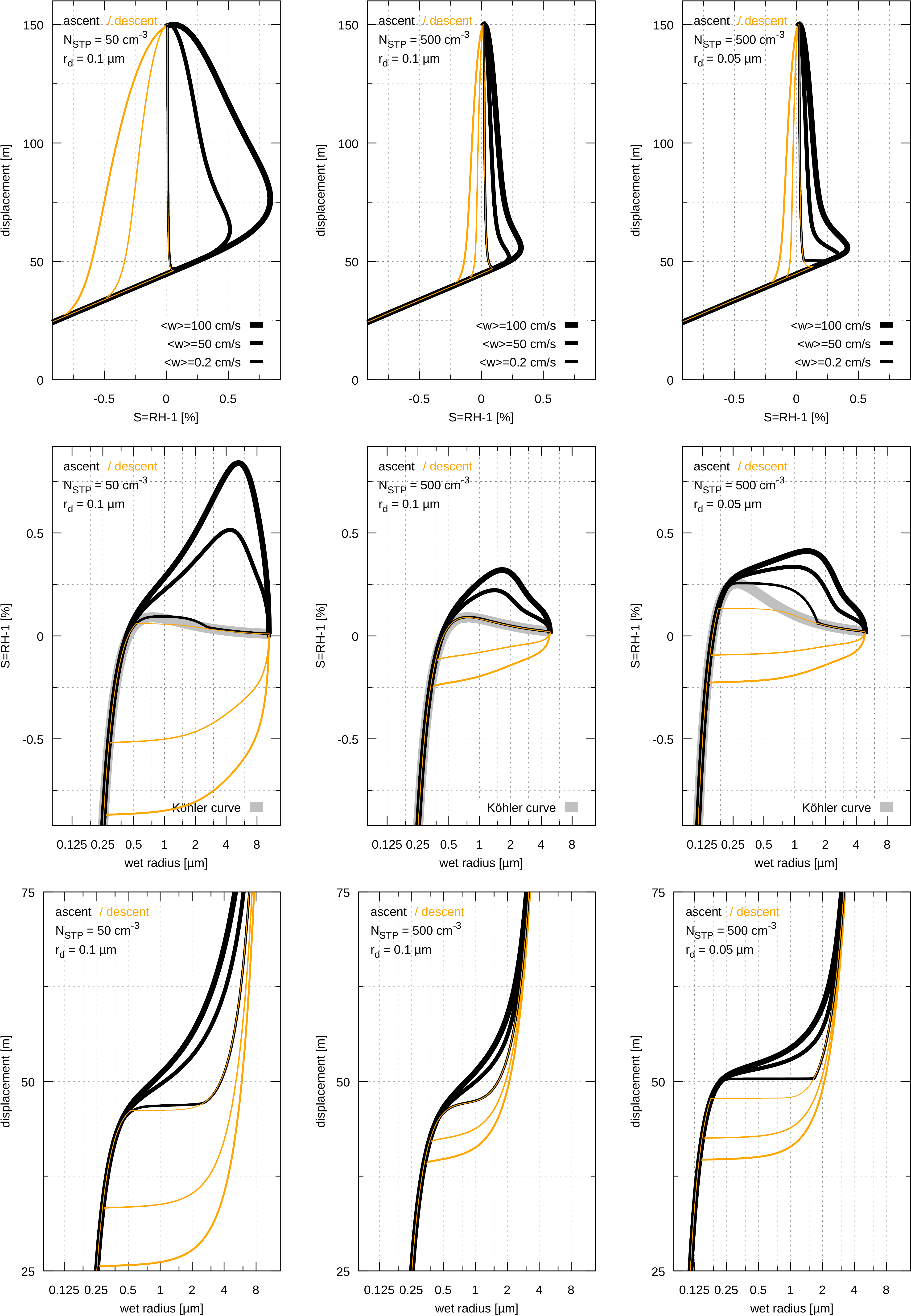}
  \caption{\label{fig:rhloop}
    Results of numerical simulations discussed in section~\ref{sec:rhloop}.
  }
\end{figure*}

Figure~\ref{fig:rhloop} summarises results of nine simulations in three types of
  coordinates: 
  displacement vs. supersaturation (the top row),  
  supersaturation vs. wet radius (the middle row, same coordinates as in Fig.~\ref{fig:koehler}) and
  displacement vs. wet radius (bottom row).
The nine model runs correspond to three sets of aerosol parameters 
  (left, middle and right columns) and three values of mean vertical velocity
  (depicted by line thickness).
The varied aerosol input parameters are the concentration ($N_\text{STP}$ of 
  50 and 500 \unit{cm^{-3}}, STP subscript corresponding to the values at 
  standard temperature and pressure)
  and the dry radius ($r_\text{d}$ of 0.1 and 0.05 \unit{\mu m}).
In all panels, black lines correspond to air-parcel ascent (activation)
  and orange lines correspond to the descent (deactivation).
Besides integration results, the panels in the middle row feature the K\"ohler
  curve plotted with thick grey line in the background.

The plots depict that for mean velocities of $100$~\unit{cm/s} and $50$~\unit{cm/s}
  activation and deactivation are not symmetric and happen far from equilibrium
  (the K\"ohler curve).
This type of hysteresis corresponds to the kinetic limitations on the transfer
  of water molecules to/from the droplet surface what
  prevents the droplets from attaining equilibrium under rapidly changing ambient
  conditions.

At much lower velocity of $0.2$~\unit{cm/s}, the processes are symmetric and 
  match the equilibrium curve, but only
  for the $N=500$~\unit{cm^{-3}} and $r_\text{d}=0.1$~\unit{\mu m} (middle column).
A twofold decrease of the dry radius (right column) 
  as well as a tenfold decrease of particle concentration (left column) 
  both cause the system to exhibit a hysteretic
  behaviour also at the lowest considered velocity.
This hysteresis is characterised by a ``jump'' in the wet radius that qualitatively
  matches the envisioned catastrophic behaviour associated with the cusp bifurcation.
This behaviour is robust to further reduction in the vertical velocity (not shown)
  confirming a close-to-equilibrium r\'egime was attained.

The adaptive-timestep solver statistics (not shown) reveal that regardless of
  the chosen relative accuracy, for all considered input parameters, 
  there are two instants for which the solver needs to significantly reduce
  the timestep: when resolving the supersaturation maximum during activation
  and when resolving the ``jump'' back to equilibrium during deactivation.
It is a robust feature that deactivation 
  requires roughly an order of magnitude shorter timestep as compared 
  to activation 
  (ca. 0.01 s vs. 0.1 s for a relative accuracy of 10\textsuperscript{-6}).
The only exception from this rule is the symmetric case which does not feature
  the (catastrophic) ``jump'' back onto the equilibrium curve.
This confirms that it is the hysteretic behaviour that imposes the tightest 
  constraints on the timestep.

\vspace{-1.6em}
\section{Monodisperse system: limitations and applicability}
\vspace{-.5em}

The key advantage of the embraced monodisperse simulation 
  is simplicity -- in terms of model formulation, result analysis
  but also integration.
Simulations of the particle size spectrum evolution during activation are prone
  to numerical difficulties both due to the stiffness of the system
  as well as due to the sensitivity to the size spectrum discretisation 
  \citep{Arabas_and_Pawlowska_2011}.

The key inherent limitations for applicability of monodisperse simulations are
  the lack of description of the cloud droplet size spectrum shape
  and the lack of differentiation between activated and unactivated CCN.
The fact that all considered particles activate also implies that the model
  is unable to resolve the noise-induced excitations to which the system
  is susceptible.
The system exhibits an excitable behaviour if subject to fluctuations in the 
  forcing terms \citep[e.g. in the cooling rate $\dot{T}$, see][discussion 
  of Fig.~10-11 and other studies referenced therein]{Hammer_2015}.
The excitations
  influence the partitioning between activated and unactivated CCN, and 
  decay when the characteristic timescale (period) of fluctuations
  is largely longer or shorter than the activation timescale 
  discussed in section~\ref{sec:timescale}.

These limitations certainly restrain the relevance of the presented
  calculations to real-world problems.
Yet, let us underline that both the monodisperse spectrum and even
  the no-RH-coupling assumption are in fact contemporarily used in atmospheric modelling
  in the recently popularised particle-based (Lagrangian, super-droplet) 
  techniques for representing aerosol, cloud and precipitation particles
  in models of atmospheric flows 
  \citep[see ][and works referred therein]{Shima_et_al_2009,Arabas_and_Shima_2013}.
In these models, in the spirit of the particle-in-cell approach, the liquid 
  water is represented with computational particles, each representing a 
  multiplicity of real-world particles with monodisperse size.
In such models, the particles undergo repeated activation-deactivation cycles.
Consequently, the close-to-equilibrium catastrophic hysteresis observed in 
  the presented simulations, even if of no foreseeable relevance to the
  macroscopic behaviour of the large-scale cloud systems modelled with the 
  particle-based techniques, has to be taken into account 
  when developing numerical integration schemes.

\vspace{-1.6em}
\conclusions[Concluding remarks] 
\vspace{-.5em}

With this note we intend to bring attention to the presence of 
  nonlinear peculiarities in the equations governing CCN activation and
  deactivation, namely a saddle-node bifurcation and a cusp catastrophe. 
We have shown that conceptualisation of the process in terms of bifurcation
  analysis yields a simple yet practically-applicable description of the system
  allowing analytic estimation of the timescale of activation.
Both through weakly-nonlinear analysis and through numerical integration,
  we have depicted the presence of a cusp catastrophe in the system
  and the corresponding hysteretic behaviour near equilibrium (i.e., 
  for very small air-parcel velocities).
The near-equilibrium hysteresis was observed to determine the timestepping 
  constraints for numerical integration when simulating an activation-deactivation 
  cycle of a monodisperse droplet population.
It is a finding of interest for cloud modelling community since
  monodisperse activation/deactivation models of the studied type 
  play a constituting role in the more-and-more widespread 
  particle-based models of aerosol-cloud interactions.

\vspace{-1em}
\begin{acknowledgements}
We thank Hanna Pawłowska for her comments to the initial version of the manuscript.
SA acknowledges support of the Poland's National Science Centre 
  (Narodowe Centrum Nauki) [decision no. 2012/06/M/ST10/00434].
This research was supported by JSPS KAKENHI Grant-in-Aid for
  Scientific Research(B): (Proposal number: 26286089),
  and by the Center for Cooperative Work on Computational Science,
  University of Hyogo.
This study was carried out during a research visit of SA to Japan
  supported by the University of Hyogo.
\end{acknowledgements}

\bibliographystyle{copernicus}
\bibliography{notes}

\end{document}